# Scheduling rules to minimize total tardiness in a parallel machine problem with setup and calendar constraints

Jacques LAMOTHE[2]*, Francois MARMIER[2], Matthieu DUPUY[1,2], Paul GABORIT[2],
Lionel DUPONT[2]
1: Pierre Fabre S.A., Direction de la Logistique
8, rue Christian d'Espic  81106 CASTRES cedex – France
Matthieu.dupuy@pierre-fabre.com
2: Toulouse University, Mines Albi, Campus Jarlard - Route de Teillet
81013 ALBI CT cedex 09 – France
{jacques.lamothe, francois.marmier}@mines-albi.fr
* Corresponding author

**Abstract:** Quality control lead times are one of most significant causes of loss of time in the pharmaceutical and cosmetics industries. This is partly due to the organization of laboratories that feature parallel multipurpose machines for chromatographic analyses. The testing process requires long setup times and operators are needed to launch the process. The various controls are non-preemptive and are characterized by a release date, a due date and available routings. These quality processes lead to significant delays, and we therefore evaluate the total tardiness criterion. Previous heuristics were defined for the total tardiness criterion, parallel machines, and setup such as ATC (Apparent Tardiness Cost) and ATCS (ATC with setups). We propose new rules and a simulated annealing procedure in order to minimize total tardiness.

## 1   Introduction

The impact of pharmaceutical and cosmetics products on the human body requires that high levels of quality control are practiced in these industries. The quality control laboratory analyses samples from all the products batches at various stages of production: raw materials, in-process and finished products. Therefore, quality controls represent about 30% of manufacturing lead-time. The main consequence of this amount of spent time on quality is the high level of tardiness: in the industrial case examined here, about 40% of jobs were late at the beginning of the study. Long, variable and uncontrolled quality control lead times are mainly caused by the use of high technology machines such as high performance liquid chromatography (HPLC) and gas chromatography (GC) chains. This paper focuses on the scheduling problem of these machines.

During a chromatographic test, the substance to be analyzed is merged into an inert fluid flow (mobile phase) and pushed throughout a column full of a support (stationary phase) at stable and preset conditions of pressure, temperature and flow speed. This process separates the various components of the substance, which can then be quantified using detectors.

The column is chosen depending on the nature of the substance to analyze (HPLC chains), or components to look for (GC chains). A long setup is needed (several hours) every time a column is changed. An operator is sometimes required only to launch the setup. There are enough operators for all possible simultaneous setups, but they are available only during specific time-windows. The rest of the setup time is the delay required before obtaining the desired stable pressure and temperature conditions. The solution to avoid these setups delays is to group tests that use the same column.

The laboratory is composed of a set of chains and a limited number of columns. Chains differ in terms of the technology used (HPLC or GC), the admissible columns, access conditions and detectors available. This means that only part of a chain is available to perform any given test.

The plant makes hundreds different substances. A substance is analyzed at different steps of the manufacturing process. Each analysis is a job that has specific due dates and requires different chromatographic tests. Chromatographic tests are operations that can be done in parallel on different chains.

To decrease lateness, and therefore improve organization, due dates have to be met. There are two principal ways of achieving this: grouping tests that need similar columns; and launching setups during the operator's availability time-window which allows maximizes use of the time-windows. This scheduling problem appears to be a parallel machine tardiness minimization problem with specific constraints: (i) only some of the parallel machines are eligible for each operation; (ii) setups are family sequence dependent; (iii) setup must start in time windows; (iv) secondary resources (columns) are limited in number.

In this paper we will detail an approach based on a simulated annealing which allows us to minimize total tardiness. The remainder of this paper is organized as follows: in the next section we discuss how similar problems are managed according to the literature. We then propose a model and develop a resolution approach. Finally, we discuss the different results obtained.

## 2 Related research on parallel machine problems

There is considerable literature on parallel machine scheduling. Several dedicated reviews exist [1], [2], [3]. Minimization of total tardiness is an important and very common topic of industrial parallel machine scheduling problems. Since the problem $1||\Sigma T_i$ has been shown as NP-hard [4] a wide range of industrial studies has, logically, interested researchers. However, problems on machines with limited flexibility, setup and secondary resource constraints were poorly studied [5]. Our bibliographic analysis distinguishes these three characteristics.

### *2.1 Machine flexibility constraint*

Depending on the authors, when only a subset of machines can realize a given job, the machines are qualified as machine eligibility restriction [6] or multipurpose[7]. Several heuristics take this particularity into account. Pinedo (1995) shows the interest of flexibility-oriented heuristics such as LFJ or LFM-LFJ. Least Flexible Job (LFJ), which is optimal for Pm|pj=1,Mj|Cmax, gives priority to the Job which can be achieved on the smallest number of machines. LFM-LFJ first chooses the Least Flexible Machine that can achieve the smallest number of jobs and assigns to it the Least Flexible Job. However, there is no optimality proof using LFM-LFJ. Vairaktaraikis and Cai also propose two heuristics for makespan minimization by considering the flexibility of a problem without release dates: LTW-LPT and LAW-LPT [8]. The Least Total Workload – Longest Processing Time (LTW-LPT) heuristic computes the total potential load for each machine (workload of jobs which are already assigned plus workload of jobs which could be assigned). It selects the machine with the smallest potential workload and performs the job with the longest processing time. The Least Average Workload – Longest Processing Time (LAW-LPT) heuristic makes the same computation, but divides the duration of each unassigned job by the number of eligible machines. The literature shows that LAW-LPT leads to results close to the optimum and that LTW-LPT stays competitive but is not robust.

Centeno and Armacost try also to minimize the $C_{max}$ on parallel machines with eligibility restrictions [9]. They show that if the machines are not nested, the LPT/LFM rule performs better than the LFJ/LFM rule.

Considering unrelated parallel machines (job durations depend on the assigned machine) and weighted earliness and tardiness criteria, Bank and Werner compare constructive heuristics with iterative metaheuristics [10]. The constructive heuristics first sort the jobs in increasing (INC) or decreasing (DEC) order, or according to release dates (RD), largest slack (SL) or smallest slack (SS) for the machines. Using this order, the jobs are successively assigned to the machines: mindful of jobs already assigned, a job is assigned to the machine that minimizes the Cmax. Finally, on each machine, assigned jobs are rescheduled in order to minimize compromises between earliness and tardiness. The best heuristic appears to be SL-DEC. Various metaheuristics are also considered (simulated annealing and threshold accepting) with various neighborhoods (SH shift one job off machine, PI pairwise interchange, SHM shift one job on machine). Authors show that both metaheuristics are equivalent and that the SH neighborhood is recommended.

## 2.2 Setup constraint

The second particularity of the problem leads us to focus on scheduling problems with setup constraints (sequence dependent $s_{i,j}$ or not $s_i$) on parallel machines (unrelated $R_m$ or identical $P_m$). The problem being NP-hard, most authors propose metaheuristics to access good solutions[11].
Allahverdi *et al*. propose a survey of scheduling problems with setup times [12]. They classify the literature according to the environment (single machine, parallel machines, flow shops, etc.), the presence or absence of batches, the presence of setup depending (or not) on the job sequence and availability constraints. A traditional way of minimizing total tardiness is to reduce the number of setups. This naturally leads to grouping tasks into batches of jobs from the same family (following the terminology of [13]). The survey shows that few works deal with our parallel machine problem.

Liaw *et al* [14] and Rocha *et al* [15] propose branch and bound optimization for the $R_m|p_i\ s_{i,j}|\Sigma T_i$, but the complexity of the algorithm lead to huge computation times for problems of a realistic size. Webster and Azizoglu minimize the flow time, considering identical parallel machines with setup [16]. They propose two dynamic programming heuristics (forward and backward) but complexity still remains a significant factor.

Weng, Lu and Ren consider $R_m|p_i\ s_{i,j}|\Sigma w_i C_i$ (unrelated parallel machines, sequence dependent setup times and a weighted flow time criterion) [5]. They compare seven heuristics. The first six, sort the jobs according to the WSPT rule based either on the average job duration over the machines or on the minimal job duration. Then, jobs are successively assigned to the machines on the basis of either the minimal completion time, the minimal job duration, or the minimal job duration plus setup time. The seventh heuristic does not sort the jobs. It successively selects the couple (job/machine) that generates the minimum completion time. This last heuristic significantly outperforms all the others.

Lee and Pinedo [17] address the $R_m|p_i,\ s_{i,j}|\Sigma T_i$ problem. They introduce a dispatching rule ATCS as an extension of the ATC rule [18] that considers setups. It appears to give better results than WSPT and EDD. Authors also propose a simulated annealing procedure that manages a priority list of jobs and schedules them with the ATCS rule. Their neighborhood is a pairwise interchange of adjacent jobs in the list. When a job is moved to a new machine, all the following jobs on its initial machine are also moved. This enables the maintenance of machine sequences with good setup times. The job with the largest setup time is selected to define the neighbor. Park *et al*. extend this approach using neural networks to fix parameters of the ATCS heuristic [19]. Eom *et al*. propose a three phase algorithm: first groups of jobs are made that have similar due dates; secondly, for each group, jobs are sequenced in a list according to the ATCS rule and interchanges of jobs are made using a tabu search algorithm; thirdly, jobs are assigned to machines. This heuristics gives better results than ATCS [20].

Kim *et al* (2002) studied a simulated annealing procedure for the same problem. The authors propose six rules for building neighbors: interchange and insertion of lot or item, lot merge and lot split. The probability of choosing an item or lot depends on the tardiness of the item or the lot. The experiment shows the value of combining the various rules [21]. Kim *et al*. consider a batch scheduling problem with unrelated parallel machines and batch dependent setups [22]. They compare three greedy heuristics (WSPT, WEDD, a two-level batch scheduling called TH) with a simulated annealing procedure (SA). TH first prioritizes batches of jobs according to their weighed due dates, then it assigns jobs to machines in order to meet due dates and minimize setups. Experiments show that TH outperforms WSPT and WEDD rules. However, SA remains better.

Low also proposes various types of neighborhood generation for a multi-stage flow-shop scheduling problem on unrelated parallel machines with setup constraint [23]. The algorithm uses two job interchange mechanisms at each iteration. The first mechanism randomly selects two jobs, which are not processed on the same machine. The second one corresponds to the choice of two adjacent jobs processed on the same machine. The authors show that this algorithm outperforms a basic simulated annealing.

Akkiraju *et al* propose an agent based system that schedules jobs on multiple non-identical machines with multiple constraints and objectives [24]. Agents apply various heuristics to build or improve solutions. Their collaboration is based on the sharing of a population of good solutions. Authors report that their systems significantly outperform any single algorithm.

## 2.3 Resource availability constraints

Resources generally present two types of availability constraints: the first is due to the number of resources which limit the number of parallel tasks, and the second is linked to a calendar which limits the presence of the resource.

When a setup time includes detaching a die and attaching of another one, with a limited number per type of die, the literature names the constraint as secondary resource [25]. Scheduling unrelated parallel machines with secondary resource constraints is poorly studied. Tamaki *et al.* propose a mathematical programming formulation of the problem and a binary representation of admissible schedules. Their best results are obtained with a binary simulated annealing. Chen and Wu (2006) present a combination of tabu search and threshold accepting algorithm for the tardiness minimization [26]. The algorithm identifies sets of operations that use the same die on the same machine. A set comprises either all the operations, or a sub-group, or one operation. All their neighborhood mechanisms move a set from the machine with the maximum tardiness to a second machine. Some also move a set from the second machine to a third machine.

A time window constraint signifies that a date which is associated to an operation (start or end date) must take place in a given time window or a given set of time windows. In this paper, the time window constraint is expressed exclusively as the setup start date. To our knowledge, no paper takes this type of constraint into account.

## 2.4 Proposition

No rule exists in the reviewed literature that takes into account all the specific constraints of the problem. Nevertheless, some List Treatment Algorithms (LTA) have been defined and tried in closed situations.
- EDD and SPT are good basic rules used in various parallel machine problems [6];
- the First Freed Machine (FFM), ATC [18] and ATCS [17] enabled prioritization of operations in unrelated parallel machine problems with due dates and setups;
- LFM/LFO rules were evaluated for flexible parallel machine problems [6].

Metaheuristics gather a class of methods, which are able to solve a large amount of combinatorial optimization problems. Therefore, it is usually relevant to use a metaheuristic to solve scheduling problems with high complexity. Koulamas *et al.* present the simulated annealing (SA) algorithm as a general optimization technique for solving combinatorial optimization problems [2]. They underline four classes of problems that can well solved by SA: routing problems, layout problems, planning problems and scheduling problems. In the survey, many authors ([10], [17], [21], [22], [23], [25]) based their approach to find solutions on an SA.

Based on the traditional rules, we propose here to adapt the most efficient one to our problem. Based on the results obtained with the LTA, and on the potential improvement given by a SA, we propose to develop a neighborhood search starting from the LTA result as an initial solution. As the literature suggests, several neighborhood will be introduced to take advantage of the problem characteristics.

In the next part, we propose a model for our scheduling problem. The LTA algorithms will then be studied. and n SA approach will be proposed and detailed.

## 3 Model and notations

The quality service is an environment composed of *M* machines working in parallel. Not every machine is able to perform each analysis. The analysis of a product sample is a job *j*. Each sample analysis needs one or several chromatographic tests. These tests are the job's operations *i*. A sample can be split in smaller ones if necessary and operations can be achieved in parallel. Therefore routings are free. If an operation is processed on a machine, which was previously processing an operation from a different family, a setup (launched by an operator) is required. A job is ended when all its operations are complete and all jobs have been processed. To detail the proposed model of this industrial problem let introduce some notations.

### 3.1 Jobs and machines

Job characteristics are modeled as follows: for each job *j* ($j = 1...v$):
- $r_j$: *release date of job* j;
- $d_j$: *due date of job* j,

Each job *j* is composed of a set of operations $O_{i,j}$ ($i = 1...w_j$). Each operation can only be performed by a subset of the machines but its duration does not depend on the chosen machine:
- $ME_{i,j}$: *subset of the machines that are eligible for operation* $O_{ij}$;
- $p_{i,j}$: *duration of the* $i^{th}$ *operation of job* j.

To be performed, an operation requires a secondary resource $Sr_s$ ($s = 1...S$), a column, which is to be fixed on the machine. The type of column used to achieve an operation depends on the *family* of the test performed. Operations belonging to same family use the same column type in the same operating conditions (temperature, pressure, flow speed). No column change is required within a sequence of analysis made under identical conditions. Thus, two operations sequenced successively on the same machine do not require any setup if they belong to the same family. Therefore, let:

- $F_{i,j}$: *be the family of the* $i^{th}$ *operation of job* j, *which is also the type of secondary resource used for the operation;*
- $NbSr_{i,j}$: *be the secondary resource number of type* $F_{i,j}$.

In the case of a column exchange, a column equilibrium (setup) $s_{i,j}$ is required. During this setup, an operator is required to fix a column, to program the chain and to launch the operations. There are sufficient operators to launch several setups simultaneously. But they are only available during a set of time windows. The setup duration is lengthy comparing to the operating duration. So, let us consider that:

- Aop: *availability time windows of operator for setups*;
- $s_{ij}$: *setup duration of the* $i^{th}$ *operation of job* j.

## 3.2 Constraints

Four constraints are considered in this problem.
- Each operation has to be assigned only once, to only one machine and one column, respecting the limited number of columns and machines.
- Operations must be performed on eligible machines.
- A change of column on a machine leads to a lengthy setup.
- The beginning of any required setups must happen during an operator's availability time window.

## 3.3 Objective

The quality controls and the different constraints linked to its process induce significant delays, without any added value. Within a sequence of operations, tardiness is generated by waiting for a specific column, respecting the human resources timetabling constraint, changing the product family too frequently. Therefore, the criterion to minimize is the total tardiness of jobs.

$$\min \sum_{j=1}^{y} T_j$$

According to Graham's notation, the problem can be formulated as:
Pm|$r_j$, $d_j$, $p_j$, $M_j$, columns, calendar synchronisation, $s_{i,j}$, $F_j$|$\Sigma T_j$.

## 4 Resolution approach

The workshop manager has to assign and schedule a job set to a resource set (equipment) according to operators' time window constraints. To solve this problem, we propose a static algorithm composed of a list treatment algorithm completed by a neighborhood search algorithm to improve the solution obtained.

## 4.1 The list treatment algorithm

In a list treatment algorithm, at each loop, one operation is assigned and scheduled to a machine. In order to detail our algorithm, let introduce some more notations:

- $t_{i,j}^m$: starting time of operation $O_{i,j}$ if it is assigned to the machine. If $O_{i,j}$ requires a setup: $t_{i,j}^m$ is the starting time of the setup;
- $c_{i,j}^m$: completion time of operation $O_{i,j}$ in machine $m$;
- $t_m$: availability date of machine m;
- $L_m$: *the list of operations that could be scheduled on machine m at date* $t_m$. *At initialization,* $L_m = \{O_{i,j} / m \in ME_{i,j}\}$;
- $LAW_m$: *average potential load of machine* m. *It sums up the task durations that could be scheduled on* m *as suggested by [8].* $LAW_m = t_m + \Sigma$ ($p_{i,j}$ / card ($ME_{ij}$) for $O_{i,j} \in L_m$);
- $F_m$: *family of the last operation assigned to the machine* m;
- $A_s$: *availability time windows of the secondary resource of type* s.

The main algorithm:
Initialization: for all m, $L_m = \{O_{i,j} / m \in ME_{i,j}\}$, $t_m = 0$ and $F_m = \emptyset$;
1/ For all machines *m* and for all $O_{i,j} \in L_m$
  i. If $F_{i,j} \neq F_m$ Then
  $$t_{i,j}^m = \min(t / t \in [\max(t_m, r_j); +\infty[ \cap A_{op} \text{ and } [t; t + s_{i,j} + p_{i,j}] \subset A_{F_{i,j}});$$
  $$c_{i,j}^m = t_{i,j}^m + s_{i,j} + p_{i,j};$$
  ii. Else
  $$t_{i,j}^m = \min(t / [t_m; t + p_{i,j}] \subset [\max(t_m, r_j); +\infty[ \cap A_{F_{i,j}});$$
  $$c_{i,j}^m = t_{i,j}^m + p_{i,j};$$
2/ choose one couple {*m'*; $O_{k,l} \in L_{m'}$} with RULE;
3/ Assign $O_{k,l}$ to *m'*
  $t_{m'} = c_{k,l}$;
  $F_{m'} = F_{k,l}$;
  $A_{F_{k,l}} \leftarrow A_{F_{k,l}} \setminus [t_{k,l}^{m'}; c_{k,l}^{m'}[$;
For all $m \in ME_{k,l}$,
  $LAW_m = t_m + \Sigma$ ($p_{i,j}$ / card ($ME_{ij}$) for $O_{i,j} \in L_m$)
  $L_m \leftarrow L_m \setminus \{O_{k,l}\}$;
4/ If all $L_m$ are empty then END, else go to 1/;

The algorithm starts with the initialization of the list of admissible assignments for each machine. At step 1/ (earliest) starting and (earliest) completion times are computed for each machine *m* and each admissible assignment of $L_m$. Two cases are possible:
  i. The operation requires setup (and therefore an operator);
  ii. The operation does not require any setup (and can be processed automatically).

At step 2/ one assignment (*m'* ; $O_{kl}$) is selected with a RULE. RULE is a priority rule that selects an admissible assignment. The different rules applied, from the literature, are the following: EDD, ATC, ATCS, LFM, LFO, etc.

Finally, the assignment is effective at step 3/

Additionally, we propose the following rule: we propose to add two new exponential terms to ATC. The $OEE_{i,j,m}$ value measures the Overall Equipment Effectiveness due to an assignment. It measures the time of added value of an operation versus the time consumed. Therefore, it generalizes the second exponential term of the ATCS rule.

The $Fl_{i,j}$ term of the ATCOEEF rule (4), evaluates the flexibility of an operation as the number of machines on which this operation can be performed.

$$ATC(i,j,m) = \frac{1}{p_{i,j}} e^{-\frac{Atc_{i,j,m}}{k_1 \bar{p}}}, \text{ and } Atc_{i,j,m} = \max(d_i - p_{i,j} - t_m; 0) \tag{1}$$

$$ATCS(i,j,m) = \frac{1}{p_{i,j}} e^{-\frac{Atc_{i,j,m}}{k_1 \bar{p}}} e^{\frac{s_{i,j}}{k_2 \bar{s}}} \tag{2}$$

$$ATCOEE(i,j,m) = \frac{1}{p_{i,j}} e^{-\frac{Atc_{i,j,m}}{k_1 \bar{p}}} e^{\frac{OEE_{i,j,m}}{k_2}} \text{ with } OEE_{i,j,m} = p_{i,j}/(c_{i,j}^m - t_m). \tag{3}$$

$$ATCOEEF(i,j,m) = \frac{1}{p_{i,j}} e^{-\frac{Atc_{i,j,m}}{k_1 \bar{p}}} e^{\frac{OEE_{i,j,m}}{k_2}} e^{-\frac{Fl_{i,j}}{k_3}} \text{ and } Fl_{i,j} = \frac{cardME_{i,j}}{cardME} \tag{4}$$

One difficulty is defining values for the parameters $k_1$, $k_2$ and $k_3$. Let us compare two affectations $(O_{i,j}; m)$ and $(O_{k,l}; m')$ with the same duration ($p_{i,j} = p_{k,l}$).

Then, $ATCOEE(i,j,m) = ATCOEE(k,l,m) \Leftrightarrow Atc_{k,l,m'} - Atc_{i,j,m} = \bar{p}\frac{k_1}{k_2}(OEE_{k,l,m'} - OEE_{i,j,m})$

Therefore, $\bar{p}\frac{k_1}{k_2}$ expresses a period of time over which an in advance operation (Atc >0) is searched for in order to improve the machine's OEE (OEE close to 1).
In the same way,

$ATCOEE(i,j,m) = ATCOEE(k,l,m) \Leftrightarrow Atc_{k,l,m'} - Atc_{i,j,m} = \bar{p}\frac{k_1}{k_2}(OEE_{k,l,m'} - OEE_{i,j,m}) - \frac{k_1}{k_3}\frac{(cardME_{k,l} - cardME_{i,j})}{cardME}$.

Consequently, $k_1/k_3$ expresses a period of time during which an operation can be anticipated if it has a low flexibility.

### *4.2 Simulated annealing with complex neighborhoods*

Based on the initial solution, we propose to improve the quality of the solution obtained with the previously presented heuristic, by using a neighborhood search. We propose different neighbor structures to evaluate the ability to quickly find a good solution. Two types of parameters are listed to define our dedicated neighbors:
- Two structural parameters, which define what the neighbor is;
- Three parameters of generation, which define precise rules to obtain a neighbor.

### 4.2.1 Structural parameters

The first structural parameter corresponds to the "movement types". From the literature review, it is clear that two major "movement types" are used to define a neighbor: insertion or exchange [21,22,23]. Exchange tries to maintain the load on the machines while insertion is used to change the load. Following these two techniques, a selected item is either inserted before or exchanged with a second selected item. In our study, we propose to apply movements to two types of items:
- A single operation;
- A set of operations, which is a set of operations of a same family continuously sequenced on a machine without any setup or operator waiting. A pack may comprise one operation only.

The second structural parameter is the "family" constraint for choosing this second item. Two values are used: compulsory (the first and the second items must have the same family), not necessary (no constraint).

### 4.2.2 Generation parameters

The parameters of generation state the neighborhood mechanism. The literature review [21] underlines that the choice of the first item should depend on its tardiness. Therefore, we propose to use selection probability equal to its tardiness (item tardiness / total scheduling tardiness).

The choice of the machine on which the second item is looked for is important. Several cases are considered for the "machine" parameter: *idem* (same machine as the first item), EM (all the eligible machines of the first item), unif (one of the eligible machines of the first item is chosen with a uniform probability).

Considering any one machine, the second item is looked for between the first item ready date and the first item actual starting time. Between these two dates, several possibilities remain. The values of the "date" parameter can be unif (one of the possible items is uniformly selected) and late (the latest one is selected).

### 4.2.3 Complex neighborhood structure

By combining all the parameters, we can define a complex neighborhood structure. But are complex neighborhoods necessary? To answer to this question, we propose investigating three neighborhood structures:
- Simple: the only neighbor mechanism is an operation insertion with a uniform selection of "machine" and "date" (see Table 3, neighbor 0);
- Operation: the "moved items" is always a single operation. Three mechanisms (see Table 3, neighbors 1 to 3) are considered as combinations of "move types", "machine" and "family" parameter values;
- Operation-pack: "moved items" are "pack". As for Operation, the same three combinations of "move types", "machine" and "family" parameter values are considered (see table 3, neighbors 4 to 6). At the 7th place the neighborhood mechanism corresponds to an "adjacent pack interchange" on a machine.

The "simple" neighborhood structure only considers one neighbor generation mechanism, while the "operation" and "operation-pack" structures respectively consider 3 and 4 neighbor generation mechanisms. The mechanism is chosen randomly. For any given "Operation" mechanism there is an equal probability of being chosen.

| Neighbor | move types | moved items | family | machine | date | neighbor | move types | moved items | family | machine | date |
|---|---|---|---|---|---|---|---|---|---|---|---|
| 0 | insert | op | no | unif | unif | 4 | insert | pack | yes | EM | unif |
| 1 | insert | op | yes | EM | unif | 5 | exchange | pack | yes | EM | unif |
| 2 | exchange | op | yes | EM | unif | 6 | insert | pack | no | unif | unif |
| 3 | insert | op | no | unif | unif | 7 | exchange | pack | no | idem | late |

Table 1: Definition of the complex neighborhood structure

# 5  Industrial data application

In order to evaluate the performance of the proposed approach, and thanks to our industrial partner, we based our tests on an industrial application. Different data sets are stochastically derived from the industrial data to test the proposed algorithms.

## 5.1  Tests generation

The industrial laboratory studied comprises 10 machines, 20 columns, and is open for operators Monday to Friday from 8 am to 6 pm. Machines are free at the beginning of the simulation.

2 load situations are considered with respectively 70 jobs (normal load situation) and 140 jobs (high load situation). For each job, two routing cases are studied to test the workshop flexibility: 10 different routings (or successions of operations) and 20 different routings. Each routing contains 1, 2 or 3 operations (the number of operations is uniformly chosen).

Each operation has a total duration (setup time and analytical time per sample) uniformly chosen between 120 and 1440 minutes. The setup is either 50% or 75% of the total operation duration.

Job release dates are generated between -8/+5 days around the beginning of the simulation and therefore some jobs can be available at the beginning of the simulation. The due date corresponds to the release date plus a lead-time generated by a Gauss curve centered on 10 days with a standard deviation of one day.

In order to measure the impact of machine flexibility, the number of eligible machines for a routing operation is generated with a mean selected in the range "2, 4, 6, and 10" and with a standard deviation of 0.5 (except when mean equals 10, where all machines are always eligible). The number of equivalent columns to the 10% most used column is defined as being 3, then 2 for the next 30% and 1 for the rest.

For each load situation (70 or 140 jobs), we test both rooting configurations (10 and 20 different routings), two setup durations ($s_{ij}/(s_{ij}+p_{ij})$ = 50% or 75%), and four machine flexibilities (2, 4, 6 or 10). So, for each load instance, we generate 16 problem types. Each problem type is solved 10 times in order to have an average effect on the values. Therefore, in both load situations, 160 problems are solved.

## 5.2  Rules comparison

First we analyze the effect of the rules used in the list treatment algorithm. In this experiment we compare First Freed Machine (FFM)- Random (that randomly chooses an operation), FFM-ATC, FFM-ATCS, LFM-LFO, FFM-EDD, FFM-ATCOEE and FFM- ATCOEEF. In order to apply a variance analysis to validate the effects of parameters of the experiment, a check must first be made to ensure that the distribution of tardiness follows a normal distribution. Figure 1 depicts the density function of the tardiness measure for various rules. Shapes can be considered as normal. But, note that the tardiness-axis is logarithmic. Consequently, "log(Tardiness)" must be analyzed instead of "Tardiness".

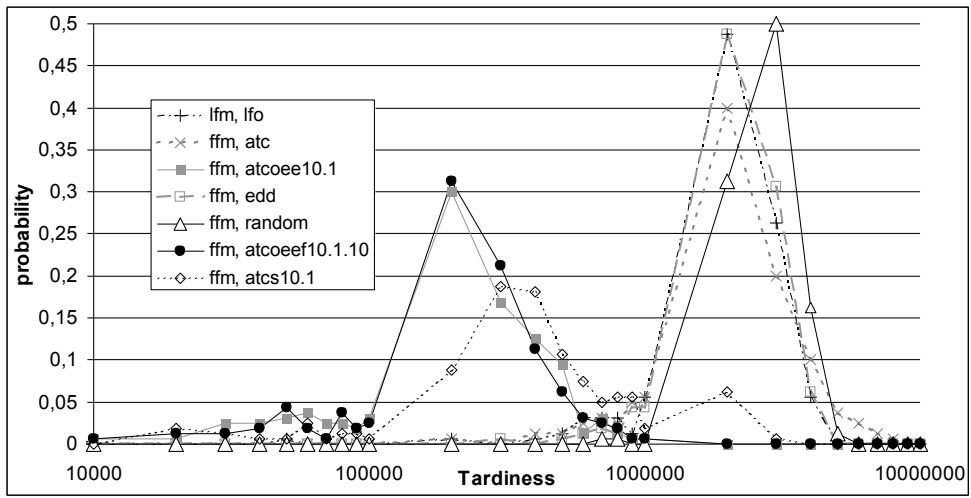

*Figure 1: density function of tardiness for various rules and 140 jobs*

Consequently, in the following tables, if the effect on log(tardiness) is $x$ when fixing a parameter $p$ to value $v$, this requires that the mean tardiness of all the experiments must be multiplied by $10^x$ to obtain the tardiness when parameter $p$ takes the value $v$.

Tables 1-2 give the results of the variance analysis for both easy (70 jobs) and difficult (140 jobs) problems. Table 1 shows that the most important effect is due to rule. Other parameters have non-significant or much less significant effects. No interaction between parameters is significant in the 70 jobs experiments (Table 2). For the 140 jobs experiment, Table 2 shows that the interaction between the rule and the number of eligible machines (Card ME) must be studied: the maximal interaction is 0.23 compared to the 0.56 of the effect of rule.

|  | 70 jobs problems | | | | 140 jobs problems | | | |
|---|---|---|---|---|---|---|---|---|
| parameter | Max effect | experimental Fisher | theoretical Fisher | Significant | Max Effect | experimental Fisher | theoretical Fisher | Significant |
| rule | 1.9 | 65.37 | 2.56 | Yes | 0.56 | 256.22 | 2.56 | Yes |
| Nb routing | 0.18 | 33.89 | 7.17 | Yes | 0.01 | 4.22 | 7.17 | No |
| sij/(sij+pij) | 0.17 | 29.96 | 7.17 | Yes | 0.07 | 93.98 | 7.17 | Yes |
| card_ME | 0.13 | 2.29 | 4.2 | No | 0.06 | 28.45 | 4.2 | Yes |

*Table 2: F-test analysis of the effect of parameters on log(tardiness)*

|  |  | 70 job problem | | | | 140 job problem | | | |
|---|---|---|---|---|---|---|---|---|---|
| parameter 1 | parameter 2 | Max Interaction | experimental Fisher | theoretical Fisher | Significant | Max Interaction | experimental Fisher | theoretical Fisher | Significant |
| rule | card_ME | 0.36 | 0.73 | 2.1 | No | 0.23 | 3.29 | 2.1 | Yes |
| sij/(sij+pij) | rule | 0.27 | 0.99 | 2.56 | No | 0.11 | 4.19 | 2.56 | Yes |
|  | rule | 0.19 | 0.97 | 2.56 | No | 0.04 | 0.82 | 2.56 | No |
| Nb routing | card_ME | 0.11 | 1.91 | 4.2 | No | 0.08 | 22.34 | 4.2 | Yes |
|  | card_ME | 0.08 | 1.15 | 4.2 | No | 0.05 | 6.43 | 4.2 | Yes |
| sij/(sij+pij) | Nb routing | 0 | 0 | 7.17 | No | 0.02 | 10.54 | 7.17 | Yes |

*Table 3: F-test analysis of the interactions between parameters on log(tardiness)*

Table 3 details the effect of the various scheduling rules. The least heuristics appears to be "random". The rule based on the measure of the flexibility (LFM/LFO) is very poor. ATC and EDD are quite similar. The ATCS rule gives interesting results in high load situations (140 jobs problem). Finally, ATCOEE and ATCOEEF rules give the best results. Considering the parameter k1, k2 and k3 used to specify ATCS, ATCOEE and ATCOEEF rules, rules are named in Table 3 using the following notation: rule.k1.k2.k3. Choosing a low value for k1 gives slightly better results in the 70 jobs situations but significantly less good results in the 140 jobs problem. Choosing a high value of k3 is always preferable. This tends to minimize the interest of the flexibility term in ATCOEEF. Consequently, one can conclude that this flexibility term is useless and that the ATCOEE rule is preferable.

| rule | 70 job problem | 140 job problem | rule | 70 job problem | 140 job problem |
|---|---|---|---|---|---|
| atcoee.10.1 | -0.7 | -0.56 | atcoee.1.1 | -0.86 | -0.15 |
| atcoeef.10.1.10 | -0.68 | -0.55 | atcoeef.1.1.10 | -0.85 | -0.15 |
| atcoeef.10.1.1 | -0.57 | -0.49 | atcoeef.1.1.1 | -0.73 | -0.09 |
| atcs.10.1 | 0.11 | -0.27 | atcs.1.1 | -0.42 | 0.05 |
| lfm_lfo | 1.52 | 0.4 | atc | 0.63 | 0.45 |
| random | 1.9 | 0.56 | edd | 0.48 | 0.45 |

*Table 4: effect of the scheduling rules on log(tardiness)*

| rule | Card ME | Interaction | rule | Card ME | Interaction | rule | Card ME | Interaction |
|---|---|---|---|---|---|---|---|---|
| atc | 2 | -0.16 | atcs.1.1 | 2 | -0.067 | atcoeef.1.1.1 | 2 | 0.111 |
|  | 4 | -0.077 |  | 4 | 0.005 |  | 4 | 0.02 |
|  | 6 | 0.003 |  | 6 | 0.011 |  | 6 | -0.029 |
|  | 10 | 0.234 |  | 10 | 0.052 |  | 10 | -0.103 |
| atcoee.1.1 | 2 | 0.011 | atcs.10.1 | 2 | -0.096 | atcoeef.1.1.10 | 2 | 0.026 |
|  | 4 | 0.031 |  | 4 | -0.004 |  | 4 | 0.013 |
|  | 6 | -0.004 |  | 6 | 0.034 |  | 6 | 0. |
|  | 10 | -0.038 |  | 10 | 0.066 |  | 10 | -0.039 |
| atcoee.10.1 | 2 | 0.105 | edd | 2 | -0.025 | atcoeef.10.1.10 | 2 | 0.128 |
|  | 4 | 0. |  | 4 | 0.017 |  | 4 | -0.015 |
|  | 6 | -0.013 |  | 6 | -0.003 |  | 6 | -0.009 |
|  | 10 | -0.092 |  | 10 | 0.011 |  | 10 | -0.104 |
| random | 2 | 0.054 | lfm_lfo | 2 | -0.122 | atcoeef.10.1.1 | 2 | 0.163 |
|  | 4 | -0.008 |  | 4 | 0.014 |  | 4 | 0.043 |
|  | 6 | -0.029 |  | 6 | 0.056 |  | 6 | -0.047 |
|  | 10 | -0.016 |  | 10 | 0.052 |  | 10 | -0.159 |

*Table 5: interaction of the scheduling rule with the number of eligible machines per operation (Card ME) for the 140 jobs problems*

Let us now consider Table 4, which depicts the effect of the interaction between the number of eligible machines and the scheduling rules when considering high load problems (140 jobs). It can be noticed that improving the workshop flexibility (improving the number of eligible machines per operation) enables tardiness to decrease when using ATCOEE and ATCOEEF rules, but surprisingly induces an increase of tardiness when using other rules (ATC, ATCS,

EDD, LFM-LFO). This latter effect can be explained by the fact that ATC, ATCS, EDD and LFM-LFO rules only take into account part of the constraints that impact decisions on operational assignment. Increasing flexibility can then induce inefficient assignments.

## 5.3 Experiments with simulated annealing

Experiments using Simulated Annealing procedures (SA) are performed using the same data and situations as for the heuristics. First, the SA starts with a descent phase on 100 iterations. Thus, the mean tardiness variation of the neighborhood structure is calculated. Then, the initial temperature of SA is evaluated so that the initial acceptance probability is fixed at 0.8 for a neighbor that generates a mean tardiness variation. Each temperature level is changed after 400 iterations or 80 accepted neighbors. The temperature is decreased according to a geometric law by a factor of 0.95. The SA is finished when three temperature levels are reached without any acceptation, or after 15000 iterations. This can be defined as a "quick" SA resolution. In order to measure the worth of a longer resolution, we also report the results when applying a SA procedure, called 'OP+PA SA 0.98', where 0.98 is used for the geometrical law and no limit exists on the number of iterations.

This experiment also aims at studying the worth of a complex neighborhood structure. So, for the "quick" SA resolution, three neighborhood structures are compared as suggested in §4.2.3:
- 'SIMPLE SA 0.95' is a basic SA algorithm that only uses operation insertion mechanisms;
- 'OP SA 0.95' only moves operations but with various neighborhood mechanisms;
- 'OP+PA SA 0.95' applies the various types of movement to operations and packs;

Finally, results are also compared to a list treatment algorithm using the ATCOEE rule.

Tables 5 and 6 report the results of the variance analysis of the effects and interactions of the experiment parameters for both situations (70 and 140 job problems). Here too it can be seen that the major effect is due to the type of SA algorithm. No significant or important interaction can be seen in both situations. This means that the best algorithm is the best for all the situations experimented.

| parameter | 70 job problem | | | | 140 job problem | | | |
|---|---|---|---|---|---|---|---|---|
| | Max effect | experimental Fisher | theoretical Fisher | Significant | Max effect | experimental Fisher | theoretical Fisher | Significant |
| Algo | 0.949 | 22.021 | 3.720 | Yes | 1.489 | 180.272 | 3.720 | Yes |
| Nb routing | 0.245 | 21.477 | 7.171 | Yes | 0.075 | 3.340 | 7.171 | No |
| sij/(sij+pij) | 0.062 | 1.363 | 7.171 | No | 0.518 | 158.390 | 7.171 | Yes |
| card_ME | 0.060 | 0.161 | 4.199 | No | 0.161 | 1.759 | 4.199 | No |

*Table 6: F-test analysis of the effect of parameters on log(tardiness) when using SA algorithms*

| parameter 1 | parameter 2 | 70 job problem | | | | 140 job problem | | | |
|---|---|---|---|---|---|---|---|---|---|
| | | Max Interaction | experimental Fisher | theoretical Fisher | Significant | Max Interaction | experimental Fisher | theoretical Fisher | Significant |
| Algo | card_ME | 0.176 | 0.138 | 2.562 | No | 0.295 | 1.220 | 2.562 | No |
| sij/(sij+pij) | Algo | 0.127 | 0.735 | 3.720 | No | 0.399 | 12.546 | 3.720 | Yes |
| Nb routing | Algo | 0.061 | 0.120 | 3.720 | No | 0.059 | 0.203 | 3.720 | No |
| Nb routing | card_ME | 0.108 | 0.463 | 4.199 | No | 0.090 | 0.663 | 4.199 | No |
| sij/(sij+pij) | card_ME | 0.035 | 0.051 | 4.199 | No | 0.061 | 0.292 | 4.199 | No |
| sij/(sij+pij) | Nb routing | 0.002 | 0.001 | 7.171 | No | 0.069 | 2.844 | 7.171 | No |

Table 7: F-test analysis of the interaction of parameters on log(tardiness) when using SA algorithms

|                | 70 job problem | 140 job problem |
|----------------|---------------|-----------------|
| ATCOEE         | 0.949         | 1.489           |
| OP SA 0.95     | -0.311        | -0.562          |
| OP+PA SA 0.95  | -0.328        | -1.014          |
| OP+PA SA 0.98  | -0.347        | -1.083          |
| SIMPLE SA 0.95 | 0.038         | 1.17            |

Table 8: effect of the scheduling SA Algorithm on log(tardiness)

Table 7 shows the effect of the choice of an SA algorithm on log(Tardiness). Naturally, the best algorithm appears to be 'OP+PA SA 0.98', as this uses much more iteration and probably finds near-optimal solutions. But the performance of 'OP+PA SA 0.95' is almost as good. The gap with 'OP+PA SA 0.98' on log(tardiness) is 0.02, which means a Tardiness gap of $(10^{0.02}-1) = 4.4\%$ for 70 job problems and $10^{0.07}-1= 17.4\%$ for 140 job problems. As it uses far fewer iterations, 'OP+PA SA 0.95' appears to be a very good compromise.

'OP SA 0.95' also performs well with simple problems (70 job problem) but less well with harder problems: for the 140 job problem the gap with 'OP+PA SA 0.98' is 0.52 on log(tardiness), which means $10^{0.52} –1= 232\%$ on tardiness. This means that in difficult situations 'OP SA 0.95' needs more iterations to converge.

'SIMPLE SA 0.95' is the worst SA algorithm. The gap with 'OP+PA SA 0.98' is significant. In 15000 iterations, 'SIMPLE SA 0.95' remains far from convergence.

As a consequence, these results show the value of implementing the proposed combination of neighborhood mechanisms to ensure quick, good and robust convergence of the SA algorithm.

The comparison between ATCOEE, the best List Treatment Algorithm, and 'OP+PA SA 0.98', the best SA algorithm, shows that the gap is enormous and becomes all the larger as the problem becomes harder (1800% for 70 job problems and 37200% for 140 job problems). This gap shows the difference between a heuristic rule that needs less than a second of computation time, and a meta-heuristic procedure that needs much more computation time (nearly 5 minutes for 15000 iterations on a Sun Ultra 40 1GHz with 2GB of memory). From an industrial point of view, List Treatment algorithms are easy to implement in a ERP system. This gap illustrates the attraction of investing in dedicated optimization software. From an academic point of view, this gap suggests that future works are necessary to find better heuristic procedures.

# 6 Conclusion

This paper deals with a flexible parallel machine scheduling problem at an industrial pharmaceutical control laboratory. This problem refers to the resources bottleneck, the chromatography chains. Specific constraints such as calendar meeting points between operations and operators, secondary resource constraints and family dependent setups are considered.

Heuristic rules from the literature are investigated (EDD, LFM, LFO, ATC, and ATCS). Two new rules are proposed as extensions of the ATC principle: ATCOEE and ATCOEEF. They include a measurement of the added value of an assignment and of solution flexibility.

These heuristics are tested on a set of experimental data that correspond to various real-life situations. The experimental study shows that: (i) ATCOEE dominates the other heuristics in all situations; (ii) machine flexibility does not really influence heuristic behavior as good solutions are obtained even with low flexibility.

The performance of the ATCOEE rule suggests to test it in other application area. Indeed the OEE term can take into account many type of constraints met in various industrial contexts.

In order to improve the results, the proposed heuristic is complemented by a simulated annealing procedure (SA). Results show the worth of a complex neighborhood structure in terms of the quality of a solution and the convergence speed of the SA algorithm.